\documentclass[final,5p,times,twocolumn]{elsarticle}

\usepackage{amsmath,amsfonts,amssymb, bm}
\usepackage{graphicx, graphics,epsfig, subfigure, color}
\usepackage{xcolor}
\usepackage{microtype}
\usepackage{siunitx}
\usepackage[colorlinks=true,linkcolor=blue,citecolor=blue,urlcolor=blue]{hyperref}
\usepackage{caption}
\usepackage{appendix}
\usepackage{dcolumn}
\usepackage{multirow}
\usepackage{setspace}
\usepackage{threeparttable}
\usepackage{tabularx}
\usepackage{booktabs}
\usepackage{float}
\usepackage{dblfloatfix}
\usepackage{makecell}
\newcolumntype{d}{D{.}{.}{-1}}

\newcommand{\lag}{\mathcal{L}}
\newcommand{\calP}{\mathcal{P}}
\newcommand{\mpl}{M_{\rm Pl}}
\newcommand{\dd}{\mathrm{d}}

\newcommand{\ns}{n_s}

\newcommand{\beq}{\begin{equation}}
\newcommand{\eeq}{\end{equation}}
\newcommand{\ber}{\begin{eqnarray}}
\newcommand{\eer}{\end{eqnarray}}
\newcommand{\nn}{\nonumber}

\journal{Physics Letters B}

\begin{document}

\begin{frontmatter}
\title{Constraints on non-canonical chaotic inflation from ACT DR6 and BICEP/Keck data}
\author[loc1]{Wei Yang}
\ead{yw1214@nuaa.edu.cn}

\author[loc1]{Chen-Hao Wu}
\ead{chenhao\_wu@nuaa.edu.cn}

\author[loc1,loc2]{Ya-Peng Hu\corref{cor1}}
\ead{huyp@nuaa.edu.cn}
\cortext[cor1]{Corresponding author}

\address[loc1]{Center for the Cross-disciplinary Research of Space Science and Quantum-technologies (CROSS-Q), College of Physics, Nanjing University of Aeronautics and Astronautics, 29 Jiangjun Road, Nanjing City, Jiangsu Province 211106, China}
\address[loc2]{Key Laboratory of Aerospace Information Materials and Physics (NUAA), MIIT, Nanjing 211106, China}

\begin{abstract}
In this study, we precisely evaluated the feasibility of the chaotic inflation model within a non-canonical kinetic framework. By applying the slow-roll approximation and imposing constraints on the equilateral non-Gaussianity $f_{\rm NL}^{\rm equil}$, we imposed constraints on the feasible range of the potential index $n$. We established physical bounds for the non-canonical parameter $\alpha$. To obtain precise parameter constraints, we solved the primordial perturbation equations numerically and conducted a rigorous MCMC analysis by using a comprehensive joint P-ACT-LB-BK18 dataset. For these potentials $n=1/3$, $2/3$, and $1$, our results respectively tightly limit $\alpha$ to the levels of $8.8^{+1.6}_{-2.8}$, $11.7^{+1.7}_{-2.6}$, and $16.4^{+3.7}_{-7.0}$, within the corresponding $1\sigma$ confidence intervals. Meanwhile, the required number of $e$-foldings naturally converges to $N \simeq 54$, without the need for fine-tuning. These findings confirm that non-standard mechanisms can resurrect excluded chaotic inflation models within the $1\sigma$ allowed regions of high-precision cosmological data.
\end{abstract}


\end{frontmatter}

\section{Introduction}
The inflationary paradigm provides a compelling framework for the very early universe, successfully resolving the initial-condition problems of the standard Big Bang model and seeding primordial density perturbations\cite{Linde:1981mu, Guth:1980zm}. For years, the Planck observations constrained the primordial scalar spectral index to $n_s=0.9649\pm0.0042$ \cite{Planck:2019kim}, while setting stringent upper limits on the tensor-to-scalar ratio $r$\cite{BICEP:2021xfz}. Recently, the latest observations from the Atacama Cosmology Telescope (ACT DR6), combined with Planck, CMB lensing, and the Dark Energy Spectroscopic Instrument (DESI) baryon acoustic oscillation (BAO) measurements\cite{DESI:2024mwx, DESI:2024uvr}, suggest a shift in the scalar spectral towards a higher value, approximately $n_s = 0.9743\pm0.0034$ \cite{AtacamaCosmologyTelescope:2025blo}. Adding B-mode polarization data from the BICEP/Keck telescopes (BK18)\cite{BICEP:2021xfz} at the South Pole to this combined dataset (P-ACT-LB) tightens the upper limit on the tensor-to-scalar ratio to $r < 0.038$ (P-ACT-LB-BK18)\cite{AtacamaCosmologyTelescope:2025nti}. 

The advent of the latest observational data has disfavored several prominent inflationary models, including Starobinsky inflation. In response, numerous recent works have investigated diverse inflationary scenarios in light of these updated constraints. It has been widely demonstrated that models ruled out in the canonical framework can be resurrected by modifying gravity \cite{NooriGashti:2026lwk, Aldabergenov:2025kcv, Nojiri:2026hij, Yuennan:2025kde, Ketov:2025cqg, Addazi:2025qra, Odintsov:2025bmp, Zhu:2025twm, Saini:2025jlc, Ellis:2025ieh, Wolf:2025ecy, Ahmed:2025sfm, He:2025bli, Yogesh:2025wak, Odintsov:2025eiv, Keskin:2025zqq, Salvio:2025izr, Yang:2024ntt}, nonminimal couplings and Palatini realizations \cite{Bostan:2025jkt, NooriGashti:2025gug, Gao:2025viy, Gao:2025onc, Dioguardi:2025mpp, Yi:2025dms, McDonald:2025tfp, Pallis:2025nrv, Dioguardi:2025vci, Yuennan:2026fcn, Yuennan:2025mlg, Oikonomou:2025xms}. Beyond altering the gravitational framework, another well-motivated extension to standard inflation involves modifying the kinetic term of the scalar field, widely known as non-canonical or k-inflation \cite{Mukhanov:2005bu, Unnikrishnan:2008ki, Rezazadeh:2014fwa, Cespedes:2015jga, Stein:2016jja, Serish:2025ian, Gialamas:2025ofz}. 

The most distinctive kinematic feature of the non-canonical framework with the power-law kinetic is that the propagation speed of scalar perturbations (the sound speed $c_s$) can be subluminal ($c_s < 1$), and the predicted tensor-to-scalar ratio $r$ is dynamically suppressed by a geometric factor proportional to $c_s$ \cite{Unnikrishnan:2012zu, Mishra:2022ijb, Lola:2020lvk}. This mechanism provides a natural pathway to lower the amplitude of primordial gravitational waves without altering the underlying potential, thereby offering a potential rescue for simple classical models. Specifically, the classical monomial chaotic inflation potentials $V(\phi)=V_0\phi^n$ were ruled out by earlier standard observations. Previous studies demonstrated that embedding these models within a non-canonical kinetic framework could successfully resurrect specific cases, such as $n=2$ or $n=4$, bringing their predictions in line with the Planck constraints\cite{Unnikrishnan:2012zu, Mishra:2022ijb}. However, the recently updated observation by a preference for a larger scalar spectral index $n_s$ and a drastically tightened upper bound on the tensor-to-scalar ratio $r$ compels us to critically re-evaluate this model. Besides, the non-canonical framework can shift the predictions back into the currently allowed $(n_s,r)$ region, and a rigorous analysis and understanding of how to constrain the model parameters remains of crucial importance.

In this paper, we perform a comprehensive analysis of chaotic inflation models within the non-canonical framework in light of the latest observations. Our primary objective is to utilize the updated data to rigorously map out the currently surviving parameter space. We adopt a two-step approach combining analytical estimation with exact numerical inference. From the perspective of approximate analysis, we utilize the observed constraints and the equilateral non-Gaussianity bound from Planck 2018 to analytically estimate the feasible range for $n$ and establish theoretical lower bounds for the non-canonical parameter $\alpha$. In addition, we numerically solve the exact background and perturbation equations. By performing a rigorous Markov Chain Monte Carlo (MCMC) analysis utilizing the latest P-ACT-LB-BK18 joint dataset to sample the posterior distributions, the precise constraints of the model parameters were obtained.

The structure of this paper is organized as follows. In Section 2, we briefly review the theoretical framework of non-canonical inflation, including the background evolution and primordial perturbations. In Section 3, we present the analytical approximation under slow-roll conditions to evaluate the viable range for the potential index $n$ and establish theoretical bounds for $\alpha$. Section 4 details our exact numerical methodology and presents MCMC posterior constraints derived from the latest joint observational datasets. Finally, we summarize our main conclusions in Section 5. 

\section{Non-canonical framework}\label{section:2}
\subsection{The background evolution equation}
In this section, we focus on the following generic action given by
\beq
S=\int \dd^4x\sqrt{-g}\left[\frac{\mpl^2}{2}R+\lag(X,\phi)\right].
\label{eq:action}
\eeq
where $X = -\frac{1}{2} g^{\mu\nu} \partial_\mu \phi \partial_\nu \phi$ is the standard kinetic term. Working in reduced Planck units ($\mpl=(8\pi G)^{-1/2}=1$), the Lagrangian density of the non-canonical scalar field model is given by \cite{Mukhanov:2005bu, Unnikrishnan:2008ki, Unnikrishnan:2012zu}
\beq \label{eq:Lagrangian}
\lag(X,\phi)=X\left(\frac{X}{M^4}\right)^{\alpha-1}-V(\phi).
\eeq
Here $M$ has dimensions of mass and $\alpha$ is the dimensionless non-canonical parameter. When $\alpha=1$, the Lagrangian reduces to the canonical scalar field Lagrangian
${\cal L}(X,\phi) = X - V(\phi)$. The energy density and pressure of the scalar field are \cite{Unnikrishnan:2012zu}
\ber
\rho_{\phi}&=&\left(\frac{\partial{\mathcal{L}}}{\partial{X}}\right)(2X)-\mathcal{L} =(2\alpha-1)X\left(\frac{X}{M^4}\right)^{\alpha-1}+V(\phi),\nn\\
p_{\phi}&=&\mathcal{L} =X\left(\frac{X}{M^4}\right)^{\alpha-1}-V(\phi).
\label{eq:rhop}
\eer

For a spatially flat Friedmann--Lema\^itre--Robertson--Walker (FLRW) background, $\dd s^2=-\dd t^2+a^2(t)\dd x_i\dd x^i$, one has $X=\dot\phi^2/2$, and the background equations become
\ber
H^2&=&\frac{1}{3}\left[\frac{1}{2}(2\alpha-1)\dot\phi^2\left(\frac{\dot\phi^2}{2M^4}\right)^{\alpha-1}+V(\phi)\right],
\label{eq:fried1}\\
\dot H&=&-\frac{1}{2}\alpha\dot\phi^2\left(\frac{\dot\phi^2}{2M^4}\right)^{\alpha-1},
\label{eq:fried2}\\
\ddot\phi&+&\frac{3H\dot\phi}{2\alpha-1}+\left(\frac{2M^4}{\dot\phi^2}\right)^{\alpha-1}\frac{V_{,\phi}}{\alpha(2\alpha-1)}=0.
\label{eq:eom}
\eer
The Hubble slow-roll parameters are defined by
\beq
\epsilon\equiv-\frac{\dot H}{H^2}, \qquad \eta\equiv\epsilon-\frac{\dot\epsilon}{2H\epsilon},
\label{eq:slowroll}
\eeq
and inflation requires $\epsilon\ll1$ and $|\eta|\ll1$. Under these conditions Eqs.\eqref{eq:fried1} and \eqref{eq:eom} reduce to
\ber
H^2&\simeq&\frac{1}{3}V(\phi),
\label{eq:sr1}\\
\dot\phi&\simeq&-\theta\left[\left(\frac{1}{\alpha\sqrt{3}}\right)\left(\frac{\theta V(\phi)_{,\phi}}{\sqrt{V(\phi)}}\right)(2M^4)^{\alpha-1}\right]^{1/(2\alpha-1)},
\label{eq:sr2}
\eer
where $\theta=+1$ for $V(\phi)_{,\phi}>0$ and $\theta=-1$ for $V(\phi)_{,\phi}<0$. Moreover, $V(\phi)_{,\phi}=dV(\phi)/d\phi$. The corresponding potential slow-roll expressions are
\beq
\epsilon\simeq\epsilon_V=\left[\frac{1}{\alpha}\left(\frac{3M^4}{V(\phi)}\right)^{\alpha-1}\left(\frac{V(\phi)_{,\phi}}{\sqrt{2}V(\phi)}\right)^{2\alpha}\right]^{1/(2\alpha-1)},
\label{eq:epsilonV}
\eeq
\beq
\eta\simeq\left(\frac{\alpha\epsilon}{2\alpha-1}\right)\left(\frac{2V(\phi)V(\phi)_{,\phi\phi}}{V(\phi)_{,\phi}^2}-1\right),
\label{eq:etaV}
\eeq
where $V(\phi)_{,\phi\phi}=dV(\phi)_{,\phi}/d\phi$. The number of e-folds from horizon exit to the end of inflation is
\beq
N=\int_t^{t_e}H\dd t=\int_{\phi}^{\phi_e}\frac{H}{\dot\phi}\dd\phi,
\label{eq:Ndef}
\eeq
where the subscript $e$ denotes the end of inflation.

\subsection{Scalar and tensor perturbations}

The Mukhanov--Sasaki equation for scalar perturbations is \cite{Garriga:1999vw}
\beq
u_k''+\left(c_s^2k^2-\frac{z''}{z}\right)u_k=0,
\label{eq:MSscalar}
\eeq
where Mukhanov variable $u_k\equiv z\,\mathcal{R}_k$, a prime denotes differentiation with respect to conformal time $\eta=\int \dd t/a(t)$. The sound speed and the function $z$ are given by
\ber
c_s^2&\equiv& \frac{\partial\lag/\partial X}{\partial\lag/\partial X+2X\,\partial^2\lag/\partial X^2}=\frac{1}{2\alpha-1},\label{eq:cs}\\
z&\equiv& \frac{a\sqrt{\rho_\phi+p_\phi}}{c_sH}.\label{eq:zdef}
\eer
From the above equation, we find that the sound speed is a constant. Since 
$c_s<c$ and units $c \equiv 1$, the parameter $\alpha$ must strictly satisfy the condition $\alpha>1$.  Moreover, for non-canonical inflation, the equilateral non-Gaussianity parameter is~$f_{\mathrm{NL}}^{\mathrm{equil}} = -\frac{275}{486}(\alpha - 1)$\cite{Rezazadeh:2014fwa,Chen:2006nt}.
The constraint on the equilateral non-Gaussianity parameter $f_{\rm NL}^{\rm equil}$ given by Planck 2018 is $f_{\mathrm{NL}}^{\mathrm{equil}} = -26 \pm 47 \quad (68\%~\text{C.L.})$ \cite{Planck:2019kim}. Hence, we can obtain an upper bound $\alpha \leq 130$. Therefore, in our subsequent analysis, we consider the non-canonical parameter \( \alpha \) to vary within the range $1 < \alpha \leq 130$. 

The scalar power spectrum is defined by
\beq
\calP_s(k)\equiv\left(\frac{k^3}{2\pi^2}\right)|\mathcal{R}_k|^2=\left(\frac{k^3}{2\pi^2}\right)\left|\frac{u_k}{z}\right|^2.
\label{eq:PsDef}
\eeq

Deep inside the sound horizon $(c_sk\gg aH)$, the effective potential $z''/z\approx2(aH)^2$ can be safely neglected. The positive-frequency mode corresponding to the standard Bunch-Davies vacuum is given by \cite{Garriga:1999vw}
\begin{equation}
u_k \approx \frac{e^{-i c_s k \eta}}{\sqrt{2 c_s k}} \quad (c_s k \gg aH). \label{eq:uk_sub}
\end{equation}
As the universe expands, the physical wavelength of the perturbation gets stretched. After the mode crosses the sound horizon $(c_sk\ll aH)$, the $c_s^2k^2$ term becomes negligible. The non-decaying super-horizon solution of Eq. (\ref{eq:MSscalar}) is proportional to $z$, yielding
\begin{equation}
u_k \approx C_k z \quad (c_s k \ll aH), \label{eq:uk_super}
\end{equation}
The integration constant $C_k$ is determined by matching the sub-horizon solution (\ref{eq:uk_sub}) with the super-horizon solution (\ref{eq:uk_super}) at the sound horizon crossing, defined by the condition $c_s k = aH$. Performing this matching under the de Sitter approximation gives $|C_k|^2\simeq 1/(2c_skz^2)|_{c_sk=aH}$. Consequently, at the moment of sound horizon crossing $(c_sk=aH)$, substituting Eq.\eqref{eq:uk_super} in Eq.\eqref{eq:PsDef}, we find that the scalar power spectrum takes the general form \cite{Garriga:1999vw}
\begin{equation}
\mathcal{P}_s(k) \equiv \frac{H^4}{8\pi^2 c_s (\rho_\phi + p_\phi)}.\label{eqn: scalar PS}
\end{equation}
Subsequently, by applying the slow-roll approximations for our specific non-canonical model, the scalar power spectrum is determined to be
\begin{equation}
\mathcal{P}_s(k) \simeq \left( \frac{1}{72\pi^2 c_s} \right) \left\{ \left(\frac{\alpha 6^\alpha}{M^{4(\alpha-1)}}\right) \left(\frac{V(\phi)^{5\alpha-2}}{V(\phi)_{,\phi}^{2\alpha}}\right)\right\}^{\frac{1}{2\alpha-1}}. \label{eq:final_ps}
\end{equation}
The scalar spectral index $n_s$ is defined as
\beq
\ns-1\equiv\frac{\dd\ln\calP_s}{\dd\ln k}.
\label{eq:nsDef}
\eeq
The scalar spectral index is constrained by P-ACT-LB data \cite{AtacamaCosmologyTelescope:2025blo,AtacamaCosmologyTelescope:2025nti},
\beq\label{eqn: Observation ns}
n_s = 0.9743\pm0.0034.
\eeq
The corresponding equation governing the tensor perturbations
is
\beq
v_k''+\left(k^2-\frac{a''}{a}\right)v_k=0.
\label{eq:MStensor}
\eeq
The spectrum of tensor perturbations is given by the usual expression 
\beq
\calP_T(k)\equiv2\left(\frac{k^3}{2\pi^2}\right)|h_k|^2=2\left(\frac{k^3}{2\pi^2}\right)\left|\frac{v_k}{a}\right|^2,
\label{eq:PTDef}
\eeq
where $v_{_k} \equiv ah_k$, $a$ is a scale factor and $h_k$ is the amplitude of the tensor perturbation. Hence, the tensor power spectrum at horizon exit $(k=aH)$ is
\beq
\calP_T(k)= 8\left(\frac{H}{2\pi}\right)^2\simeq \frac{2V(\phi)}{3\pi^2}.
\label{eq:PTApprox}
\eeq
Finally,
\beq
r\equiv\frac{\calP_T}{\calP_s}.
\label{eq:rDef}
\eeq
The observational upper limit on the tensor-to-scalar ratio is bounded to $r < 0.038$ at the 95\% confidence level (P-ACT-LB-BK18) \cite{AtacamaCosmologyTelescope:2025nti, BICEP:2021xfz}.

\section{Analytical approximation for chaotic inflation}
For chaotic inflation with a monomial potential of the form\cite{Linde:1983gd}
\beq
V(\phi)=V_0\phi^n, n>0
\label{eq:chaotic}
\eeq
According to the end-of-inflation condition $\epsilon(\phi_e)=1$ in Eq.~(\ref{eq:epsilonV}), we can obtain
\beq
\phi_e=\left[\left(\frac{M^{4(\alpha-1)}}{\alpha}\right)\left(\frac{3}{V_0}\right)^{\alpha-1}\left(\frac{n}{\sqrt{2}}\right)^{2\alpha}\right]^{1/[\gamma(2\alpha-1)]},
\label{eq:phiend}
\eeq
where
\beq
\gamma\equiv\frac{2\alpha+n(\alpha-1)}{2\alpha-1}.
\label{eq:gamma}
\eeq
Substituting the potential function and Eq.\eqref{eq:sr2}  into Eq.\eqref{eq:Ndef}, and utilizing Eq.\eqref{eq:phiend}, we obtain the functional relationship
between $\phi(N)$ and $N$  is given by
\beq
\phi(N)=\left[\frac{n}{\alpha}\left(\frac{6M^4}{V_0}\right)^{\alpha-1}\right]^{1/[\gamma(2\alpha-1)]}\left(N\gamma+\frac{n}{2}\right)^{1/\gamma}.
\label{eq:phiN}
\eeq
Since $H$ and $c_s$ are nearly constant during slow roll, one may use $\dd/\dd\ln k\simeq-\dd/\dd N$ at sound-horizon exit. Substituting Eq.\eqref{eq:phiN} into the scalar and tensor spectrum Eqs.\eqref{eq:final_ps} and \eqref{eq:PTApprox}, we obtain the predictions for the scalar spectral index $n_s$ and the tensor-to-scalar ratio $r$ \cite{Unnikrishnan:2012zu}
\beq
\ns=1-2\left(\frac{\gamma+n}{2N\gamma+n}\right),
\label{eq:nsanalytic}
\eeq
\beq
r=\frac{1}{\sqrt{2\alpha-1}} \left(\frac{16n}{2N\gamma+n}\right).
\label{eq:ranalytic}
\eeq

To determine the viable range of the model parameters $n$ and $\alpha$, we combine the analytical expressions for the spectral index $n_s$ and the tensor-to-scalar ratio $r$ with the latest observational constraints. First, we define a parameter $\lambda \equiv (1 - n_s)/2$. From the expression for $n_s$, we have the relation
\begin{equation}
\frac{\gamma + n}{2N\gamma + n} = \lambda \implies n(1 - \lambda) = \gamma(2N\lambda - 1).
\label{eq:n_gamma_relation}
\end{equation}

Substituting the explicit form of $\gamma \equiv \frac{2\alpha + n(\alpha - 1)}{2\alpha - 1}$ into Eq.~(\ref{eq:n_gamma_relation}) and isolating $n$, we obtain $n$ as a function of $\alpha, N$, and $\lambda$
\beq
n = \frac{2\alpha(2N\lambda - 1)}{(2\alpha - 1)(1 - \lambda) - (\alpha - 1)(2N\lambda - 1)}.
\label{eq:n_formula}
\eeq
To rigorously find the extrema of $n$, we analyze its monotonicity with respect to the parameters $\alpha$ and $\lambda$. Given the physical parameter space ($N\in [50,60]$, $\alpha > 1$, and $\lambda \sim \mathcal{O}(10^{-2})$ ensuring $2N\lambda>1$), it is straightforward to verify that $\partial n / \partial \alpha < 0$ and $\partial n / \partial \lambda > 0$. This mathematically proves that $n$ is a monotonically decreasing function of $\alpha$ and a monotonically increasing function of $\lambda$

Simultaneously, the tensor-to-scalar ratio $r$ must satisfy the observational bound $r<0.038$ \cite{AtacamaCosmologyTelescope:2025nti, BICEP:2021xfz}. To establish a general bound independent of $n$, we substitute the term $n/\gamma$ extracted from Eq.~\eqref{eq:n_gamma_relation} directly into Eq.~\eqref{eq:ranalytic}, we obtain a simplified formula decoupled from $n$
\begin{equation}
r = \frac{16(2N\lambda - 1)}{(2N - 1)\sqrt{2\alpha - 1}}.
\label{eq:r_simplified}
\end{equation}
Given the observational constraint $n_s = 0.9743 \pm 0.0034$ (which translates to $\lambda \in [0.01115, 0.01455]$), we can deduce the phenomenologically viable range for $n$ under typical $e$-folding numbers $N=50$ and $N=60$.

For $N=50$, the minimum value of $n$ is reached when $\lambda$ takes its minimum $0.01115$ and $\alpha$ takes its maximum theoretical prior $130$, yielding $n_{\rm min} \approx 0.12$. Conversely, to maximize $n$, one must maximize $\lambda$ as $0.01455$ and decrease $\alpha$. However, as indicated by Eq.~\eqref{eq:r_simplified}, decreasing $\alpha$ leads to the increase of $r$. The strict limit $r<0.038$ imposes a lower bound of $\alpha >2.37$. Substituting $\alpha = 2.37$ and $\lambda = 0.01455$ into Eq.~\eqref{eq:n_formula}, we find $n_{\rm max} \approx 0.70$. Thus, the allowed range is $0.12 < n < 0.70$. For $N=60$, following an identical logical procedure, $n_{\rm min} \approx 0.41$ is calculated at $\lambda = 0.01115$ and $\alpha = 130$. For the maximum value, taking $\lambda = 0.01455$, the limit $r<0.038$ enforces $\alpha> 3.98$, which yields $n_{\rm max} \approx 1.28$. Thus, the viable range is $0.41< n < 1.28$.

This analytical result clearly indicates that, even with the introduction of non-canonical dynamics, models with the potential index $n=2$ or $n=4$ are strongly excluded by current data. However, chaotic inflation potentials with fractional exponents and linear potentials (specifically $n = 1/3, 2/3$, and $1$, which are widely discussed in axion monodromy or string theory phenomenological models \cite{Silverstein:2008sg, McAllister:2008hb, Dong:2010in}) fall comfortably within the $1\sigma$ observational constraints. This provides a solid phenomenological motivation for us to select these three specific values for in-depth exploration.

Once the index $n$ is fixed to these specific target values, $n_s$ (and thus $\lambda$) is no longer a completely free parameter, but is intrinsically determined by $N$ and $\alpha$. Therefore, to establish an analytical lower bound on the non-canonical parameter $\alpha$ for these specific models, we refer back to the original expression for $r$ in Eq.~\eqref{eq:ranalytic}. Since $r$ is independent of the absolute energy scale parameters ($M$ and $V_0$), it is primarily suppressed by the factor $1/\sqrt{2\alpha - 1}$. Requiring the theoretical prediction to satisfy $r_{\rm obs}< 0.038$ yields a necessary condition for $\alpha$
\begin{equation}
    \alpha > \frac{1}{2} \left[ \left( \frac{16n}{r_{obs}(2N\gamma + n)} \right)^2 + 1 \right].
    \label{eq:alpha_analytical_bound}
\end{equation}
Since $\gamma$ also depends on $\alpha$ via its definition, Eq.~\eqref{eq:alpha_analytical_bound} is an implicit inequality that can be solved numerically. Crucially, as derived in our preceding analysis, the phenomenologically viable values of $n$ strongly depend on the choice of the $e$-folding number $N$. Therefore, we must evaluate these specific models at their respective favored $N$ values.

For $n=1$ and $n=2/3$, both values fall comfortably within the viable range ($0.41<n < 1.28$) for $N=60$. Evaluating Eq.~\eqref{eq:alpha_analytical_bound} at $N=60$, the standard canonical linear potential ($n=1, \alpha=1$) yields an excluded value of $r \approx 0.066$. To geometrically suppress this excessive tensor production to satisfy $r<0.038$, the non-canonical parameter is strictly required to be $\alpha>2.86$. Similarly, for the $n=2/3$ fractional potential, restoring phenomenological viability necessitates a suppression of $\alpha>1.46$. For $n=1/3$, this $\alpha$ needs to be required to fall within the range where $\alpha$ is greater than 1.

\begin{figure}[h]
    \centering
    \includegraphics[width=0.45\textwidth]{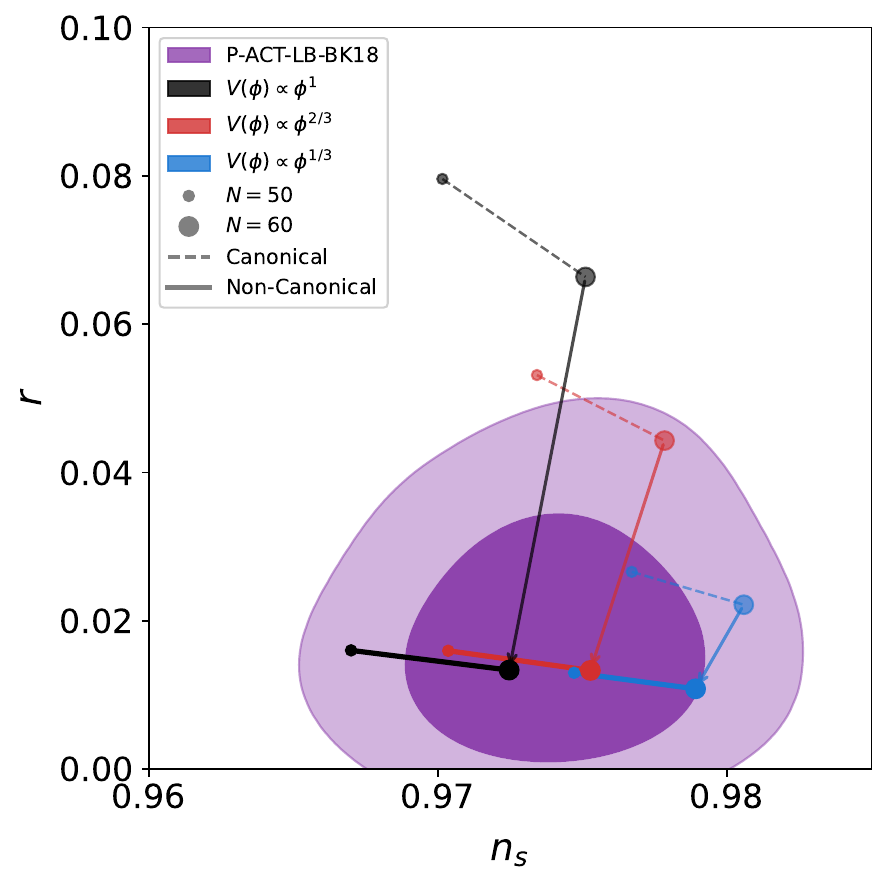}
    \caption{Observational constraints on $n_s-r$ for the
    chaotic inflation at Non-canonical framework at $k_*=0.05 Mpc^{-1}$. The $1\sigma$ and $2\sigma$ contours are the observational results from the P-ACT-LB-BK18 data \cite{AtacamaCosmologyTelescope:2025blo, AtacamaCosmologyTelescope:2025nti}. Theoretical models are shown for three power-law potentials, $V(\phi)\propto \phi^1$(black), $\phi^{2/3}$ (red), and $\phi^{1/3}$ (blue). Dashed lines denote the standard canonical limits $(\alpha = 1)$, which generally predict an excess of tensor modes. Solid lines depict the corresponding non-canonical cases evaluated at selected different values of $\alpha$. Along each trajectory, the small and large solid circles mark $N = 50$ and $N = 60$, respectively. The arrows explicitly illustrate the dynamical shift induced by the non-canonical kinetic term.}
    \label{fig:nsr}
\end{figure}

To visually demonstrate this geometrical suppression mechanism, we confront our theoretical predictions with the latest P-ACT-LB-BK18 joint constraints in the $n_s$--$r$ plane, as shown in Fig.~\ref{fig:nsr}. The dashed lines denote the canonical inflation limits ($\alpha = 1$) spanning $N \in [50, 60]$. As clearly illustrated, the standard linear potential ($V \propto \phi^1$) lies entirely outside the $2\sigma$ confidence contours, while the fractional potentials ($\phi^{2/3}$ and $\phi^{1/3}$) reside marginally at the $2\sigma$ boundaries or completely outside the tightly constrained $1\sigma$ region. This confirms that these canonical chaotic inflation models struggle to comfortably accommodate the latest observation constraints.

However, as the non-canonical parameter increases beyond the analytical lower bounds derived above, the fundamental inflationary kinematics are significantly altered. Flowing along the solid non-canonical trajectories (indicated by arrows), the tensor-to-scalar ratio $r$ experiences a dramatic geometrical suppression, accompanied by a moderate dynamical shift in the scalar spectral index $n_s$. This mechanism successfully drags the theoretically predicted values from the disfavored exterior directly into the most favored $1\sigma$ allowed region.

Combined with the broad non-Gaussianity constraint ($\alpha \le 130$), this analytical estimation outlines a phenomenologically viable window for the non-canonical strength, such as $\alpha \in (2.86, 130]$ for the $n=1$ model. Nevertheless, it is crucial to recognize that this analytical derivation relies heavily on the slow-roll approximation. To pinpoint the exact required magnitude of $\alpha$ within these broad analytical windows and to eliminate slow-roll approximation errors, it is imperative to introduce the joint observational data. Consequently, we will proceed to perform a comprehensive and exact numerical MCMC parameter estimation in the next section.

\section{Observational constraints}
\label{sec:mcmc}
To obtain the observationally preferred parameter values, we numerically integrate the exact background and perturbation equations (described in Sec.\ref{section:2}) without relying on the slow-roll approximation within the non-canonical framework. Moreover, we perform a Bayesian inference using an MCMC analysis. According to Bayes' theorem, the posterior probability distribution $\mathcal{P}(\boldsymbol{\Theta}|\mathcal{D})$ of the model parameters given the observational data is constructed as
\beq
    \mathcal{P}(\boldsymbol{\Theta} | \mathcal{D}) = \frac{\mathcal{L}(\mathcal{D} | \boldsymbol{\Theta}) \pi(\boldsymbol{\Theta})}{\mathcal{P}(\mathcal{D})},
    \label{eq:bayes}
\eeq
where $\mathcal{L}(\mathcal{D}|\mathbf{\Theta})$ denotes the likelihood function, $\pi(\mathbf{\Theta})$ is the prior distribution, and $\mathcal{P}(\mathcal{D})$ serves as the normalization evidence. We utilize the official posterior estimation chains publicly released by the ACT DR6 collaboration~ \cite{AtacamaCosmologyTelescope:2025blo, AtacamaCosmologyTelescope:2025nti}. Specifically, to accurately capture the joint constraints on both scalar and tensor perturbations, we adopt the posterior samples derived from their most comprehensive baseline dataset: ACT DR6 combined with Planck data, CMB lensing, DESI BAO, and BICEP/Keck 2018 B-mode polarization data (hereafter referred to as the P-ACT-LB-BK18 dataset). We apply a 3-dimensional Gaussian Kernel Density Estimation to these converged MCMC samples to reconstruct a continuous, non-parametric joint likelihood function for the primordial observables: the scalar spectral index $n_s$, the tensor-to-scalar ratio $r$, and the logarithm of the primordial scalar amplitude $A_s$, such that $\mathcal{L}(\mathcal{D}|\mathbf{\Theta}) \propto \mathcal{P}(n_s, r, lnA_s)$. 
\begin{table*}[h]
\centering
\caption{Prior distributions of parameters for the model.}
\label{tab:priors}
\renewcommand{\arraystretch}{1.4} 
\begin{tabular}{l c c c}
\hline\hline
\textbf{Parameter} & \textbf{$n=1/3$} & \textbf{$n=2/3$} & \textbf{$n=1$}\\
\hline
$\alpha$ & $\mathcal{U}(1.0,\, 130]$ & $\mathcal{U}(1.0,\, 130]$ & $\mathcal{U}(1.0,\, 130]$  \\
$M$ & $\mathcal{U}(10^{-5},\, 2\times 10^{-2})$ & $\mathcal{U}(10^{-5},\, 10^{-3})$ & $\mathcal{U}(10^{-5},\, 10^{-3})$ \\
$V_0$ & $\mathcal{U}(1.2\times 10^{-10},\, 1.6\times 10^{-10})$ & $\mathcal{U}(1.2\times 10^{-10},\, 1.6\times 10^{-10})$ & $\mathcal{U}(3.0\times 10^{-11},\, 8.0\times 10^{-11})$ \\
$N$ & $\mathcal{U}[50,\, 60]$ & $\mathcal{U}[50,\, 60]$ & $\mathcal{U}[50,\, 60]$\\
\hline\hline
\end{tabular}
\end{table*}

\begin{figure}[h]
    \centering
    \includegraphics[width=0.49\textwidth]{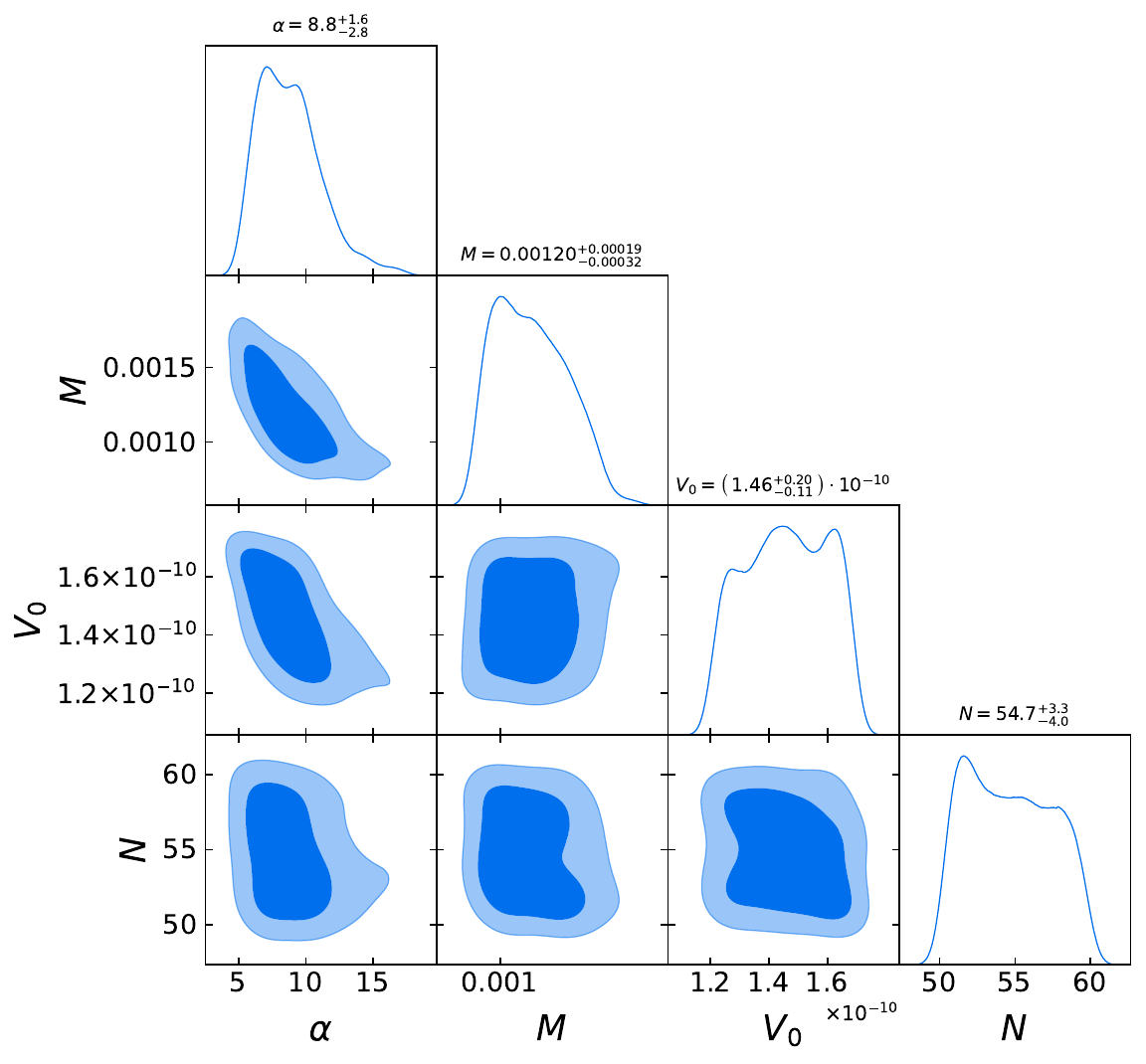}
    \caption{The corner plot for the parameters $\boldsymbol{\Theta} = \{\alpha, M, V_0, N\}$ under the potential index $n=1/3$ by using the P-ACT-LB-BK18 data \cite{AtacamaCosmologyTelescope:2025blo,AtacamaCosmologyTelescope:2025nti}.}
    \label{fig:corner13}
\end{figure}

\begin{figure}[h]
    \centering
    \includegraphics[width=0.49\textwidth]{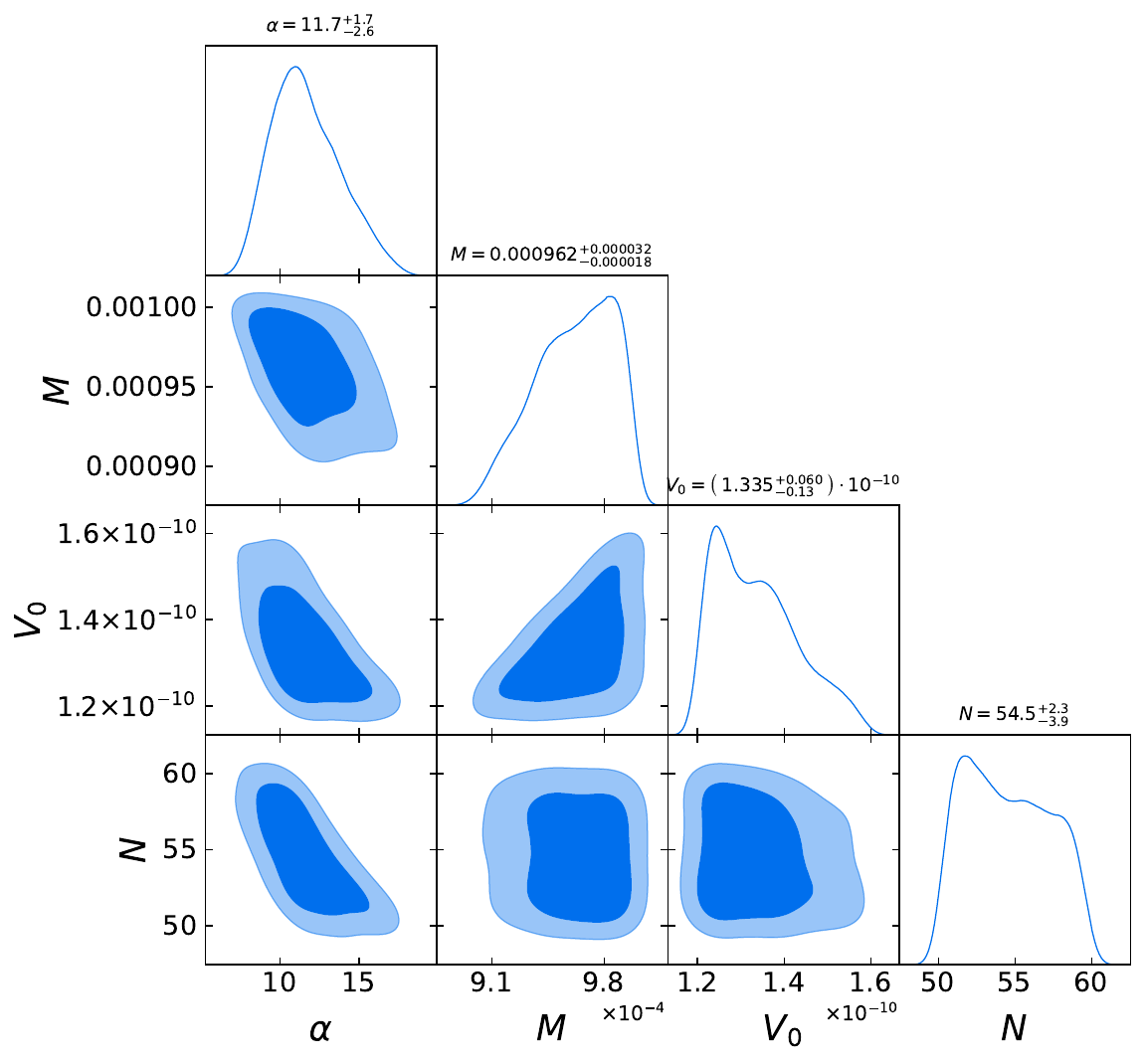}
    \caption{The corner plot for the parameters $\boldsymbol{\Theta} = \{\alpha, M, V_0, N\}$ under the potential index $n=2/3$ by using the P-ACT-LB-BK18 data \cite{AtacamaCosmologyTelescope:2025blo,AtacamaCosmologyTelescope:2025nti}.}
    \label{fig:corner23}
\end{figure}

\begin{figure}[h]
    \centering
    \includegraphics[width=0.49\textwidth]{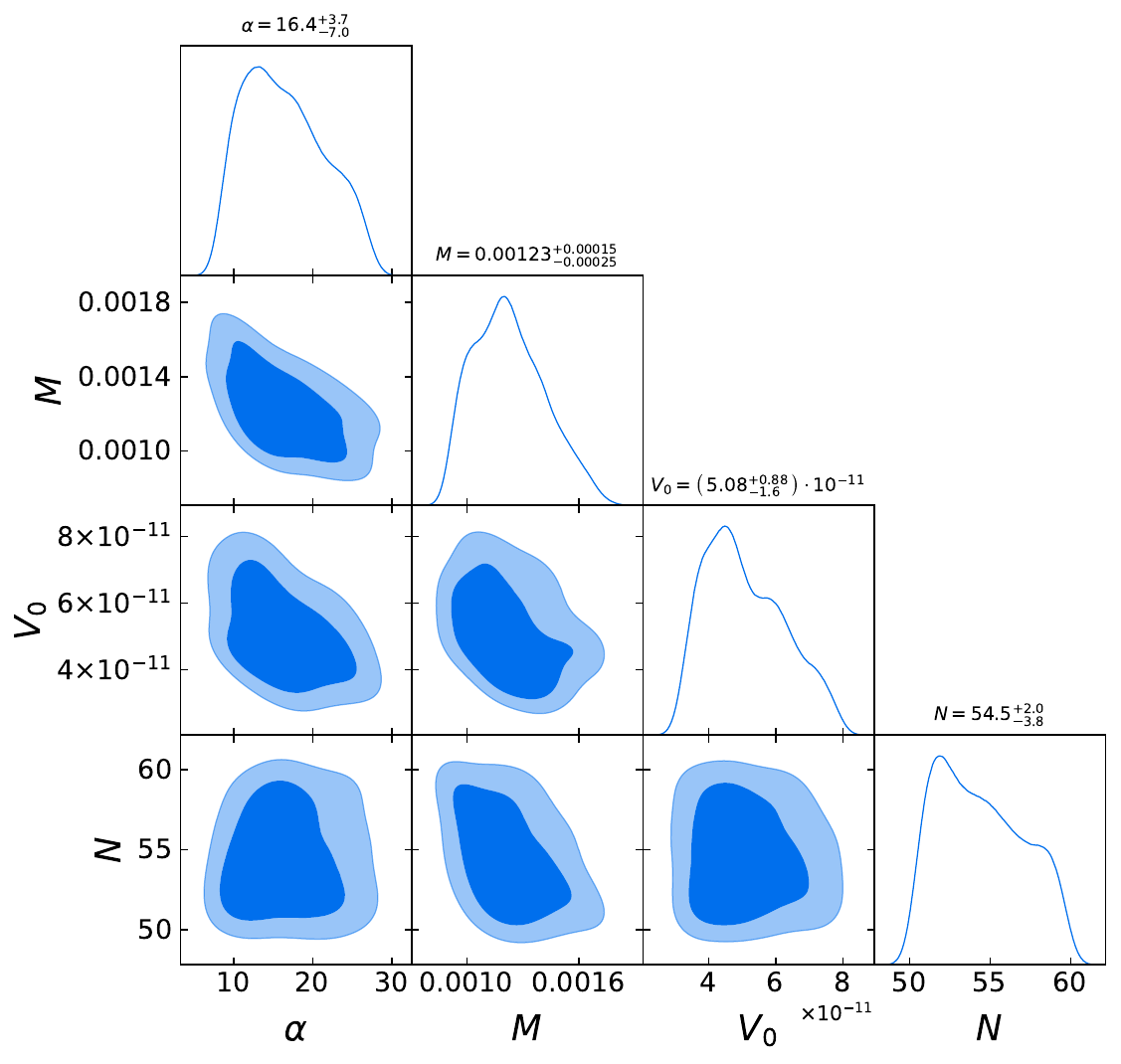}
    \caption{The corner plot for the parameters $\boldsymbol{\Theta} = \{\alpha, M, V_0, N\}$ under the potential index $n=1$ by using the P-ACT-LB-BK18 data\cite{AtacamaCosmologyTelescope:2025blo,AtacamaCosmologyTelescope:2025nti}.}
    \label{fig:corner1}
\end{figure}

To efficiently explore the posterior distributions of our model parameters $\Theta$, we employ the affine-invariant MCMC ensemble sampler \texttt{emcee}. During our numerical MCMC pipeline, the base parameter space is spanned by $\mathbf{\Theta} = \{\alpha, M, V_0, N\}$, corresponding to the non-canonical parameter, the fundamental mass scale, the potential amplitude, and the number of $e$-foldings, respectively.  We assign uniform priors $\mathcal{U}$ to all sampled parameters. For the non-canonical parameter $\alpha$, the prior range $\alpha\in\mathcal{U}(1,130]$ is strictly determined by theoretical and observational requirements. As derived under the Eq.\eqref{eq:cs}, the condition for a subluminal sound speed $(c_s<1)$ requires $\alpha>1$, while the upper bound $\alpha\leq130$ is  constrained by the Planck 2018 constraints on equilateral non-Gaussianity $f_{\text{NL}}^{\text{equil}}$\cite{Planck:2019kim}. The folding number $N$ is selected from the range $\mathcal{U} [50,60]$, and it corresponds to the required CMB window. For the mass scale $M$ and the potential amplitude $V_0$, their designated boundaries are fundamentally anchored by the strict observational constraint on the primordial scalar perturbation amplitude. The specific prior boundaries for each specific power-law index ($n=1/3,2/3$ and $1$) are detailed in Table \ref{tab:priors}. The corner plots of the posterior distributions for the base parameter space $\boldsymbol{\Theta} = \{\alpha, M, V_0, N\}$ for the three representative three power-law potentials are shown in Fig.\ref{fig:corner13}-\ref{fig:corner1}. The diagonal sub-panels show the posterior distributions with the marked $1\sigma$ constraint, while the off-diagonal panels present the joint contours. 

\begin{table*}[h]
\centering
\renewcommand{\arraystretch}{1.35}
\caption{Posterior distribution of $1\sigma$ intervals for the parameters of the non-canonical chaotic inflation.}
\label{tab:posteriors}
\begin{tabular}{l c c c}
\hline\hline
\textbf{Parameter} & \textbf{$n=1/3$} & \textbf{$n=2/3$} & \textbf{$n=1$} \\
\midrule
$\alpha$ & $8.8^{+1.6}_{-2.8}$ & $11.7^{+1.7}_{-2.6}$ & $16.4^{+3.7}_{-7.0}$ \\
$M$ & $0.00120^{+0.00019}_{-0.00032}$ & $0.000962^{+0.000032}_{-0.000018}$ & $0.00123^{+0.00015}_{-0.00025}$ \\
$V_0$ & $(1.46^{+0.20}_{-0.11})\times10^{-10}$ & $(1.335^{+0.06}_{-0.13})\times10^{-10}$ & $(5.08^{+0.88}_{-1.6})\times10^{-11}$ \\
$N$ & $54.7^{+3.3}_{-4.0}$ & $54.5^{+2.3}_{-3.9}$ & $54.5^{+2.0}_{-3.8}$ \\ 
\bottomrule
\end{tabular}
\end{table*}

\begin{figure*}[htbp]
    \centering
    \includegraphics[width=0.32\textwidth]{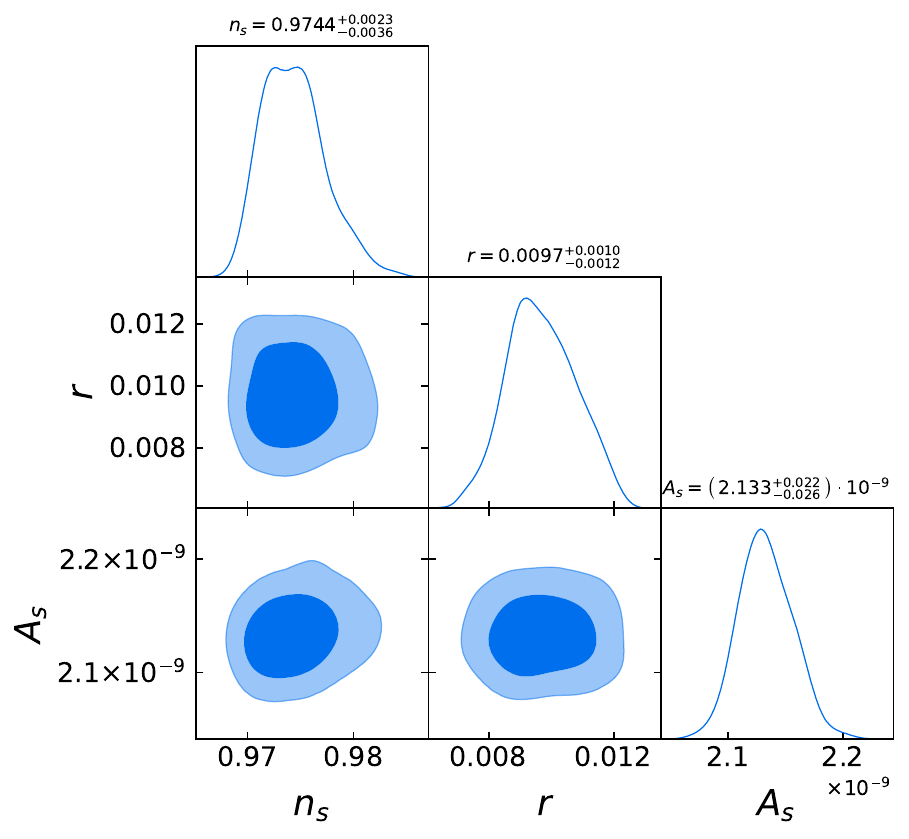}
    \includegraphics[width=0.32\textwidth]{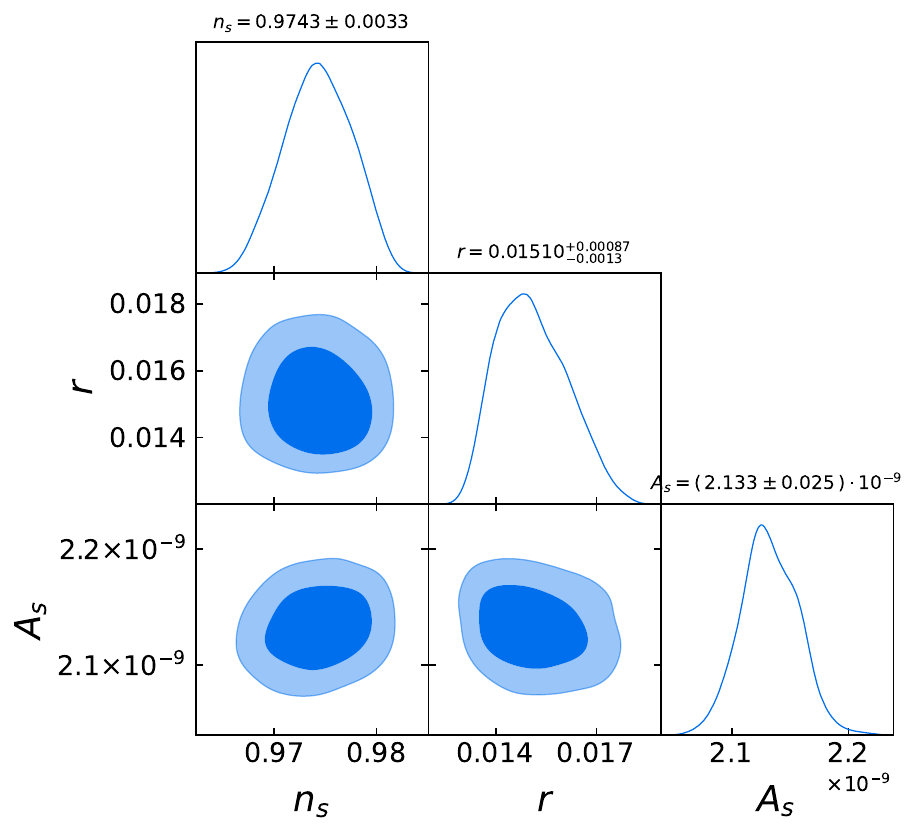}
    \includegraphics[width=0.32\textwidth]{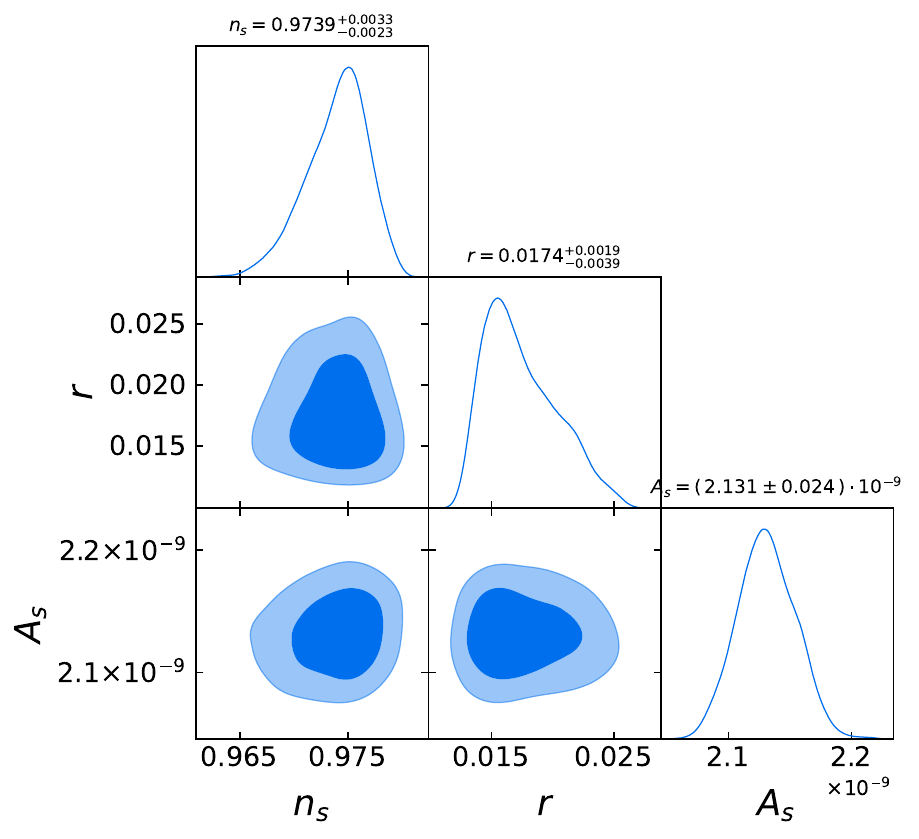}
    \caption{The corner plot for the primordial observables ($n_s$, $r$, $A_s$) derived from our MCMC analysis. From left to right, the panels correspond to the non-canonical chaotic inflation models with potentials $V \propto \phi^{1/3}$, $V \propto \phi^{2/3}$, and $V \propto \phi^{1} $, respectively. The inner and outer contours represent the $1\sigma$ and $2\sigma$ confidence levels.}
    \label{fig:observables_corner}
\end{figure*}

By analyzing these joint and marginal posteriors, we can clearly observe how the non-canonical dynamics respond to different potential indices. Specifically, the data tightly constrains the non-canonical parameter to $\alpha = 8.8^{+1.6}_{-2.8}$ for $n=1/3$ (Fig.\ref{fig:corner13}), $\alpha = 11.7^{+1.7}_{-2.6}$ for $n=2/3$ (Fig.\ref{fig:corner23}), and requires a significantly elevated value of $\alpha = 16.4^{+3.7}_{-7.0}$ for the linear potential $n=1$ (Fig.\ref{fig:corner1}), which are detailed in Table \ref{tab:posteriors}. This precisely quantifies how the strict upper bound on $r$ from BK18 dynamically forces the non-canonical parameter space to shrink towards larger values of $\alpha$ to sufficiently suppress the sound speed $c_s$. Crucially, these precise numerical inferences highlight the fundamental limitations of the analytical approximations discussed in Section 3. While our purely analytical approach, which solely required $r < 0.038$, successfully provides qualitative minimum bounds, the exact MCMC posteriors localize $\alpha$ at significantly higher central values with remarkably narrow confidence intervals, which leads to a much stronger non-canonical modification than theoretical analysis. Concurrently, to maintain the fixed amplitude of the primordial scalar power spectrum $A_s$, the overall potential scale $V_0$ decreases as $n$ and $\alpha$ increase. Notably, despite these intricate dynamics, the e-folds parameter is robustly constrained to $N \simeq 54$, ensuring that the standard thermal history of the universe is perfectly accommodated. 

To explicitly present the phenomenological results, we extract the posterior distributions for the primordial observables ($n_s$, $r$, $A_s$), as shown in Fig.~\ref{fig:observables_corner}. Within the constrained parameter space, the scalar spectral index and the amplitude of scalar perturbations peak around $n_s \approx 0.974$ and $A_s \sim 2.1 \times 10^{-9}$, respectively, which are consistent with the observations. For the tensor-to-scalar ratio, the $1\sigma$ numerical predictions are tightly constrained to $r = 0.0097^{+0.0010}_{-0.0012}$ (for $n=1/3$), $r = 0.01510^{+0.00087}_{-0.0013}$ (for $n=2/3$), and $r = 0.0174^{+0.0019}_{-0.0039}$ (for $n=1$). These results indicate that the non-canonical mechanism successfully suppresses the tensor-to-scalar ratio $r$ below the current BICEP/Keck upper bound ($r<0.038$), while maintaining it stably at the order of $\mathcal{O}(10^{-2})$, which provides a clear and specific theoretical target for future observations.

\section{Conclusions and discussion}

In this paper, we have presented a comprehensive analysis of monomial chaotic inflation within a non-canonical kinetic framework. Our primary objective was to determine whether a dynamically reduced sound speed can resurrect these theoretically well-motivated, yet observationally severely constrained, early-universe models in light of the latest precision data from P-ACT-LB-BK18.

Analytically, by synergizing the slow-roll approximation with the Planck constraints on equilateral non-Gaussianity ($f_{\rm NL}^{\rm equil}$), we mapped out the phenomenologically viable parameter space for the potential index $n$. We demonstrated that the viable parameter space restricts the index to $0.12 <n < 0.70$ for $N=50$, and $0.41 <n <1.28$ for $N=60$. This analytical derivation robustly excludes the classic steep monomial potentials ($n=2$ or $n=4$), thereby guiding our focused numerical investigation toward fractional and linear potentials ($ n=1/3, 2/3$, and $1$). Moreover, this analytical estimation provides a theoretically viable upper and lower bound for the non-canonical parameter $\alpha$. 

Beyond analytical estimations, we compute the background and exact primordial perturbation equations numerically. We then perform a rigorous MCMC analysis to place stringent constraints on the parameters ($\alpha, M, V_0, N$) using the comprehensive P-ACT-LB-BK18 joint dataset. Our numerical posteriors confirm that steeper potentials intrinsically produce larger tensor amplitudes, thus requiring a larger non-canonical parameter $\alpha$ to sufficiently suppress the sound speed. Specifically, at the $1\sigma$ region, the non-canonical parameter is constrained to $\alpha = 8.8^{+1.6}_{-2.8}$ for the flattest potential ($n=1/3$), increasing to $\alpha = 11.7^{+1.7}_{-2.6}$ for $n=2/3$, and requiring a significantly stronger modification of $\alpha = 16.4^{+3.7}_{-7.0}$ for the linear case ($n=1$). Alongside these constraints, the foundational energy scales of the potentials are precisely bounded, with the mass parameter settling at $M \sim \mathcal{O}(10^{-3})$ and the amplitude fixed at $V_0 \sim \mathcal{O}(5\times10^{-11}\sim 10^{-10})$ across all configurations.

Furthermore, the posterior distributions for the $e$-folding number consistently converge to $N \simeq 54$ across all three viable models. This naturally satisfies the standard theoretical requirement for the horizon exit without any fine-tuning of the initial conditions. Ultimately, our full MCMC analysis of the latest datasets has precisely quantified the required model parameters and further confirms that simple fractional and linear chaotic potentials can be resurrected to fully satisfy the $1\sigma$ observational constraints.

In addition, forthcoming CMB experiments such as LiteBIRD \cite{LiteBIRD:2022cnt} and CMB-S4 \cite{CMB-S4:2016ple} are expected to reach a tensor-to-scalar ratio sensitivity of $r \lesssim 0.001$. A future detection of $r \sim 0.01$ would perfectly align with our moderate non-canonical models, whereas a strict upper bound of $r \lesssim 0.001$ would require a severely large $\alpha$ that potentially exceeds the non-Gaussianity upper bound ($\alpha \leq 130$), offering a clear criterion to modify this framework. Furthermore, distinguishing our models from other scenarios that predict similarly suppressed $r$ remains an interesting issue, and primordial non-Gaussianity may be the key distinguishing signature. While standard canonical models predict a vanishingly small equilateral non-Gaussianity, the subluminal sound speed of our framework ($c_s<1$) implies a specific negative equilateral non-Gaussianity, $f_{\rm NL}^{\rm equil} = -\frac{275}{486}(\alpha - 1)$. Consequently, future joint observations of $r$ and $f_{\rm NL}^{\rm equil}$ could provide complementary clues to better understand the underlying inflationary dynamics.

\section*{Acknowledgements}
This work is supported by the National Natural Science Foundation of China (NSFC) under Grant No.12175105

\bibliographystyle{elsarticle-num} 
\biboptions{sort&compress}
\bibliography{References}

\end{document}